\documentclass[useAMS,usenatbib]{mn2e}
\usepackage{graphicx}
\usepackage{txfonts}
\usepackage{color}
\title[Galaxy clusters mass and pressure]{Mass and pressure constraints on galaxy clusters from interferometric SZ observations}
\label{firstpage}
\author[Olamaie et~al.]{Malak Olamaie,$^{1}$\thanks
{Email:mo323@mrao.cam.ac.uk} Michael P. Hobson$^{1}$ and  
Keith J. B. Grainge$^{1,2}$\\
$^{1}$ Astrophysics Group, Cavendish Laboratory, 19 J. J. 
Thomson Avenue, Cambridge, CB3 0HE\\
$^{2}$ Kavli Institute for Cosmology Cambridge, Madingley 
Road,Cambridge, CB3 0HA}

\begin{document}

\date{Accepted ??????; Received ???????}

\pagerange{\pageref{firstpage}--\pageref{lastpage}}

\pubyear{2011}

\maketitle
\begin{abstract}
Following on our previous study of an analytic parametric model to describe 
the baryonic and dark matter distributions in clusters of galaxies with 
spherical symmetry, we perform an SZ analysis of a set of simulated clusters and present their mass and pressure profiles. The simulated clusters span a wide range 
in mass, $2\times 10^{14}\rm{M_\odot}\, < M_{\rm{tot}}(r_{200})<1.0
\times 10^{15}\, {\rm M}_\odot $ , and observations with the Arcminute Microkelvin Imager (AMI) are simulated through their Sunyaev-
Zel'dovich (SZ) effect. We assume that the dark 
matter density follows a Navarro, Frenk and White (NFW) profile and that the 
gas pressure is described by a generalised NFW (GNFW) profile. By 
numerically exploring the probability distributions of the cluster 
parameters given simulated interferometric SZ data in the context of 
Bayesian methods, we investigate the capability of this model and analysis technique to 
return the simulated clusters input quantities. We show that considering 
the mass and redshift dependency of the cluster halo concentration parameter 
is crucial in obtaining an unbiased cluster mass estimate and hence deriving the 
radial profiles of the enclosed total mass and the gas pressure out to $r_{200}$.
\end{abstract}
\begin{keywords}
galaxies: clusters-- cosmology: observations -- methods: data analysis
\end{keywords}
%------------------------------------------------------------------------------%
\section{Introduction}
\label{sec:intro}
Determining the properties of clusters of galaxies such as their total 
and baryonic mass offers an independent and powerful cosmological tool  
to constrain the parameters of the $\Lambda \rm{CDM}$ 
model. The mass distribution of clusters is usually measured 
using a variety of observational methods, including X-ray, Sunyaev--
Zeldovich (SZ)(Sunyaev \& Zeldovich 1970; Birkinshaw 
1999; Calrstrom, Holder \& Reese 2002), and gravitational lensing analyses. These methods 
are often based on parameterised cluster models for the distribution 
of the cluster dark matter and the thermodynamical properties of its 
intra-cluster medium (ICM). However, these various approaches usually 
lead to different estimates of the cluster mass. This is due to either 
extrapolating to halo masses and to redshifts that are not well 
sampled by the data or fitting for model parameters to which the data are insensitive.

In this letter, we perform a detailed analysis of a sample of nine 
simulated SZ observations of galaxy clusters in a mass range of $2\times 10^{14}\rm{M_\odot}\, < M_{\rm
{tot}}(r_{200})<1.0 \times 10^{15}\, {\rm M}_\odot $ at redshift $z=0.3$ as if observed with Arcminute Microkelvin 
Imager (AMI) (AMI Consortium: Zwart et~al. 2008). To study the cluster total mass, we 
use the model described in Olamaie et~al. (2012), which has the following main 
characteristic features: (1) host halo density profile follows a 
 Navarro, Frenk and White (NFW) (Navarro et~al. 1997) profile 
and the gas pressure is described by a generalised NFW (GNFW) profile 
(Nagai et~al. 2007) with fixed shape parameters, both in accordance with 
numerical simulations; (2) the gas distribution is in hydrostatic 
equilibrium with the cluster total gravitational potential dominated by 
dark matter and both dark matter and gas are spherically symmetric; and 
(3) the local gas fraction is much less than unity throughout the 
cluster, i.e. $ \frac{\rho_{\rm gas}(r)}{\rho_{\rm tot}(r)}\ll 1$ for all 
$r$. This final assumption allows us to write $\rho_{\rm tot}(r) = \rho_
{\rm DM}(r) + \rho_{\rm gas}(r)\approx \rho_{\rm DM}(r)$. We show that 
assuming the dark matter halo concentration parameter as an independent 
free parameter in the analysis has the potential to introduce biases in the cluster 
mass estimate, and hence it is crucial to consider 
the mass and redshift dependency of this parameter in the analysis. 
Throughout, we assume a $\rm{\Lambda CDM}$ cosmology with $\, \Omega_
{\rm M}=0.3 \, , \, \Omega_{\rm \Lambda}=0.7\, , \, \sigma_{\rm 8}=0.8\, ,
\, h =0.7\, ,\, w_{\rm 0}=-1\, ,\, w_{\rm a}=0$.
\section{Modelling and analysis of simulated interferometric SZ observations}
\label{sec:analysis}
As the SZ surface brightness is proportional to the line-of-sight 
integral of the  pressure of the hot plasma in the ICM, SZ analysis of 
galaxy clusters provides a direct measurement of the pressure 
distribution of the ICM. 

The observed SZ surface brightness in the direction of electron reservoir 
may be described as
\begin{equation}\label{deltaI}
\delta I_\nu=T_{\rm CMB}yf(\nu)\frac{\partial B_\nu}{\partial T}\Big\vert_
{T=T_{\rm CMB}}.
\end{equation}
Here $B_\nu$ is the blackbody spectrum, $T_{\rm CMB}=2.73 $~K (Fixsen 
et~al. 1996) is the temperature of the CMB radiation, $f(\nu)=\left(x\frac
{e^x+1}{e^x-1}-4\right)\left(1 + \delta (x , T_{\rm e})\right)$ is the frequency 
dependence of thermal SZ signal, $x=\frac{h_{\rm p}\nu}{k_{\rm B}T_{\rm 
CMB}}$, $h_{\rm p}$ is Planck's constant, $\nu$ is the frequency and $\rm
{k_{\rm B}}$ is Boltzmann's constant. $\delta (x , T_{\rm e})$ takes into 
account the relativistic corrections due to the relativistic thermal 
electrons in the ICM and is derived by solving the Kompaneets equation up 
to the higher orders (Rephaeli~1995, Itoh et~al. 1998, Nozawa et~al. 
1998, Pointecouteau et~al. 1998 and Challinor and Lasenby 1998). It 
should be noted that at 15 GHz (AMI observing frequency) $x= 0.3$ and 
therefore the relativistic correction, as shown by Rephaeli (1995), is 
negligible for $k_{\rm B}T_{\rm e} \leq 15\, \rm{keV}$. The dimensionless 
parameter $y$, known as the Comptonization parameter, is the integral of 
the number of collisions multiplied by the mean fractional energy change 
of photons per collision, along the line of sight
\begin{equation}\label{eq:ypar}
 y = \frac{\sigma_{T}}{m_{\rm e}c^2} \int_{-\infty}^{+\infty}{n_{\rm e}
(r)k_{\rm B}T_{\rm e}(r){\rm d}l}
= \frac{\sigma_{T}}{m_{\rm e}c^2} \int_{-\infty}^{+\infty}{P_{\rm e}(r)
{\rm d}l},
\end{equation}
where $n_{\rm e}(r)$, $P_{\rm e}(r)$ and $T_{\rm e}$ are the electron 
number density, pressure and temperature at radius $r$ respectively. $
\sigma_{\rm T}$ is Thomson scattering cross-section, $m_{\rm e}$ is the 
electron mass, $c$ is the speed of light and $dl$ is the line element 
along the line of sight. It should be noted that in equation~(\ref 
{eq:ypar}) we have used the ideal gas equation of state.

Moreover, the integral of the Comptonization $y$ parameter over the solid 
angle $\Omega$ subtended by the cluster ($Y_{SZ}$) is proportional to the 
volume integral of the gas pressure. It is thus a good estimate for the 
total thermal energy content of the cluster and hence its mass (see e.g.
Bartlett \& Silk 1994). The $Y_{SZ}$ parameter in both cylindrical and 
spherical geometries may be described as
\begin{eqnarray}\label{eq:Ycylsph}
Y_{\rm cyl}(R)&=& \frac{\sigma_{T}}{m_{\rm e}c^2}\int_{-\infty}^{+\infty}
{\rm{d}l}\,\int_{0}^{R}{P_{\rm e}(r)2\pi s \, \rm {d}s}, \\
Y_{\rm sph}(r)&=& \frac{\sigma_{\rm T}}{m_{\rm e}c^2}\int_{0}^{r}{P_{\rm 
e}(r')4\pi r^{'2}\rm {d}r'},  
\end{eqnarray}
where $R$ is the projected radius of the cluster on the sky. 

In this 
context we use the model described in Olamaie et~al. (2012), with its corresponding 
assumptions on the dynamical state of the ICM, to model the SZ signal 
and determine the radial profiles of $M_{\rm {tot}}$ and $P_{\rm e}$ for nine simulated clusters. The model assumes that the the dark matter 
density follows a Navarro, Frenk and White (NFW) profile (Navarro et~al. 1997) and 
the ICM plasma pressure is described by the generalised NFW (GNFW) profile (Nagai 
et~al. 2007),
\begin{equation}\label{eq:DMdensity}
\rho_{\rm {DM}}(r)=\frac{\rho_{\rm {s}}}{\left(\frac{r}{R_{\rm s}}\right)\left(1 + \frac{r}{R_{\rm s}}
\right)^2},
\end{equation}
\begin{equation}\label{eq:GNFW}
P_{\rm e}(r)=\frac{P_{\rm {ei}}}{\left(\frac{r}{r_{\rm p}}\right)^c\left(1+\left(\frac{r}{r_{\rm
p}}\right)^{a}\right)^{(b-c)/ a}},
\end{equation}
where $\rho_{\rm {s}}$ is an overall normalisation coefficient,
$R_{\rm s}$ is the scale radius where the logarithmic slope of the
profile ${\rm d}\ln \rho(r)/{\rm d}\ln r=-2$, $P_{\rm {ei}}$ is also an 
overall normalisation coefficient of the pressure profile and $r_{\rm p}$ 
is the scale radius. It is common to define the latter in terms
of $r_{\rm 500}$, the radius at which the mean enclosed density is 500 
times the critical density at the cluster redshift, and the gas
concentration parameter, $c_{\rm 500}=r_{\rm 500}/r_{\rm p}$. The
parameters $(a,b,c)$ describe the slopes of the pressure profile at
$r\approx r_{\rm p}$, $r> r_{\rm p}$ and $r \ll r_{\rm p}$
respectively. In the simplest case, we follow Arnaud et~al. (2010) and
fix the values of the gas concentration parameter and the slopes to be
$(c_{\rm 500},a,b,c)=(1.156,1.0620, 5.4807, 0.3292)$. It is also common 
practice to define the halo concentration parameter, $c_{200}=\frac{r_
{200}}{R_{\rm s}}$, where $r_{200}$ is the radius at which the enclosed 
mean density is $200$ times the critical density at the cluster redshift. 
The cluster model parameters: $\rho_{\rm s}$, $R_{\rm s}$ 
and ${P_{\rm {ei}}}$ and hence the pressure and the integrated mass 
distributions may be derived under the following assumptions: spherical symmetry; hydrostatic 
equilibrium; and that the local gas fraction is much less than unity, 
equations (3) to (11) in Olamaie et~al. (2012). It should be noted that in 
Olamaie et~al. (2012), we considered the halo concentration parameter, $c_{200}$ as 
an input free parameter and studied the cluster profiles for a 
distribution of concentrations at a given mass and redshift. However, the results of 
our analysis showed that $c_{200}$ remains unconstrained. $c_{200}$ is a physical 
parameter, which reflects the background density of the Universe, and so is not a parameter 
that just defines the shape or the slope of the profile that can not be constrained by SZ only analysis 
of galaxy clusters.  This therefore suggests that the concentration depends on the other 
physical sampling parameters for a given set of cosmological parameters, i.e., 
mass and redshift. Hence, in this letter, we consider such a dependency in our 
analysis and instead derive $c_{200}$. 

We also note that 
The results of the studies on the relation between the structural properties of the 
dark matter halos such as concentration, spin and shape with mass and the 
redshift from both $N$- body simulations of cosmological structure 
formation in a Cold Dark Matter (CDM) Universe and observations of clusters 
of galaxies show a clear dependence of the concentration parameter on the 
halo mass and its redshift.  This is based on the fact, as first discussed  by 
Navarro, Frenk \& White (1996) and (1997), that the concentration parameter 
reflects the background density of the Universe at its formation time and 
hence as small objects form first in a hierarchical universe, lower mass 
halos will be more concentrated than the massive ones (Navarro, Frenk \& 
White 1997; Eke et~al. 2001; Bullock et~al. 2001; Pointecouteau et~al. 
2005; Macci\`{o} et~al. 2007; Vikhlinin et~al. 2006; Comerford et~al. 2007; Salvador-
Sol\'{e} et~al. 2007; Buote et~al. 2007; Neto et al. 2007; Duffy et~al. 2008; 
Mandelbaum et~al. 2008; Rudd et~al. 2008; Corless et~al. 2009; Mu\~{n}oz-Cuartas 
et~al. 2011; Ettori et~al. 2011; Bhattacharya et~al. 2011). Moreover, all of these 
studies show that the concentration-mass relation is well fitted by a power law 
over the mass range $10^{11}-10^{15}h^{-1}{\rm M}_\odot$. However, studies on 
determining the normalisation coefficient and the slope of such relation are still 
ongoing, and in particular, cluster observations have yet to investigate this further. 
In this letter, we, therefore, decided to use the relation derived by 
Neto et~al. (2007) from $N$-body simulation which has also been used by Corless 
et~al. (2009) in their weak lensing analysis of three real clusters, namely
\begin{equation}\label{eq:c200M200}
c_{200}=\frac{5.26}{1+z}\left( \frac{M_{\rm {tot}}(r_{\rm 200})}{10^{14}h^{-1}{\rm M}_\odot} \right)^
{-0.1}.
\end{equation}
We also studied the $c_{\rm {vir}}-M_{\rm {vir}}$ relation given by Mu\~{n}oz-
Cuartas et~al. (2011) where the normalization coefficient and the slope vary with 
cosmic time. However, the results were the same as using equation (\ref
{eq:c200M200}) within our cluster halo mass range. 

We generate a sample of nine simulated SZ clusters equally spaced 
in the mass range $2\times 10^{14}\rm{M_\odot}\, < M_{\rm{tot}}(r_{200})<1.0
\times 10^{15}\, {\rm M}_\odot $ using the above mentioned model, equation (\ref
{eq:c200M200}) and the input parameters $M_{\rm {tot}}(r_{\rm 200})$, $z$ and $f_{\rm 
{gas}}(r_{200})$ listed in Tab.~\ref{tab:simpars}; this set of parameters fully 
describes the Comptonization $y$ parameter. Further details of generating simulated SZ 
skies and observing them with a model AMI small array (SA) are described in Hobson \& 
Maisinger (2002), Grainge et~al.\ (2002), Feroz et~al. (2009) and AMI Consortium: 
Olamaie et~al. (2012).

\begin{table}
\caption{ $M_{\rm {tot}}(r_{\rm 200})$ used to generate simulated clusters and 
the thermal noise levels reached in the simulated observations of the cluster
sample. Here $\sigma_{\mathrm {SA}}$ refers to the thermal noise levels reached
in SA maps. All the clusters are generated at fixed redshift $z=0.3$ and fixed
$f_{\rm g}(r_{\rm 200})=0.13$. \label{tab:simpars}.}
\begin{tabular}{@{}lcc@{} }
\hline
Cluster&$M_{\rm {tot}}(r_{\rm 200})\, 10^{14}\,\rm{M_\odot}$&$\sigma_{\mathrm {SA}}$ ($\mathrm{mJy beam^{-1}}$)\\\hline
clsim1&$2.0$ & $0.05$\\
clsim2&$3.0$ & $0.06$\\
clsim3&$4.0$ & $0.07$ \\
clsim4&$5.0$ & $0.06$\\
clsim5&$6.0$ & $0.077$ \\
clsim6&$7.0$ & $0.08$\\
clsim7&$8.0$ & $0.087$\\
clsim8&$9.0$ & $0.07$\\
clsim9&$10.0$ &$ 0.067$\\  \hline
\end{tabular}
\end{table}

The sampling parameters in our Bayesian analysis are $\mbox{\boldmath$\Theta$}_{\rm 
c}\equiv (x_{\rm c}, y_{\rm c}, M_{\rm {tot}}(r_{\rm 200}),f_{\rm g}(r_{\rm 200}), 
z)$, where $x_{\rm c}$ and $y_{\rm c}$ are cluster projected position on the sky. We 
further assume that the priors on sampling parameters are separable (Feroz et~al.\ 
2009) such that
\begin{equation}\label{eq:prior}
 \pi(\mbox{\boldmath$\Theta$}_{\rm c})=\pi(x_{\rm c})\,\pi(y_{\rm c})\,\pi(M_T
(r_{\rm 200}))\,\pi(f_{\rm g}(r_{\rm 200}))\,\pi(z).
\end{equation}
We use Gaussian priors on cluster position parameters, centred on the pointing 
centre and with standard-deviation of 1 arcmin and adopt a $\delta$ function prior 
on redshift $z$. The prior on $M_{\rm {tot}}(r_{\rm 200})$ is taken to be uniform in 
log$M$ in the range $M_{\rm {min}} = 10^{14}\,\rm{M_ \odot}$ to $M_{\rm 
{max}} = 6\times10^{15}\, \rm{M_\odot}$ and the prior of $f_{\rm {gas}}(r_{\rm 
200})$ is set to be a Gaussian centred at the  $f_{\rm {gas}}=0.13$ with a width of 
$0.02$ (Vikhlinin et~al. 2005, 2006; Komatsu et~al. 2011; Larson et~al. 2011). It 
should be noted that for the two low mass clusters we set the minimum mass to $M_
{\rm {min}} = 0.4\times10^{14}\,\rm{M_ \odot}$ in the prior range. A summary of the priors and their ranges are presented in Tab.~ \ref{tab:clpriors2}.

\begin{figure}
\includegraphics[width=80mm]{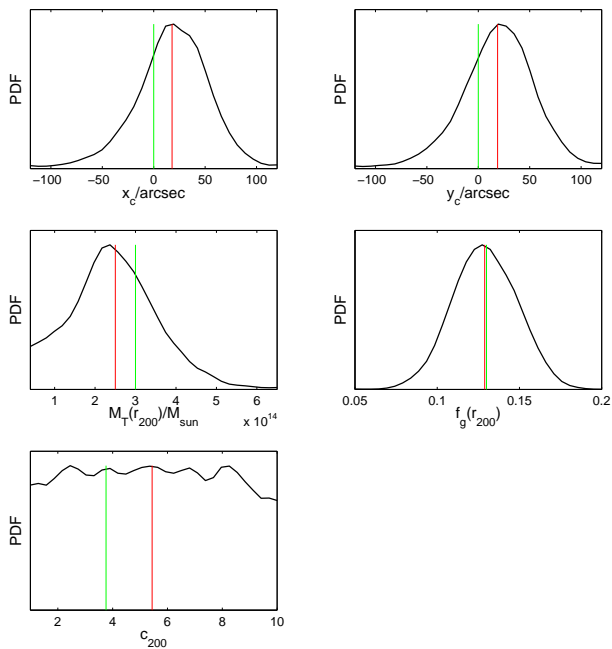}
\caption{1D marginalised posterior distributions of sampling parameters of clsim2 when 
$c_{200}$ is also assumed to be an input parameter. Green vertical lines are the true 
cluster parameter values as given in Tab.~\ref{tab:simpars} and the red vertical lines 
are the mean of the probability distributions of the parameters.} \label{fig:varycpos}
\end{figure}

\begin{figure}
\includegraphics[width=80mm]{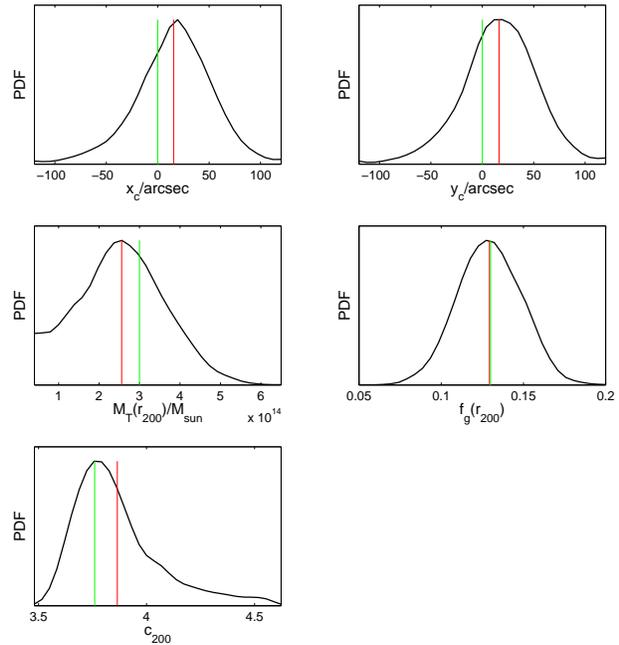}
\caption{1D marginalised posterior distributions of sampling parameters of clsim2 when 
$c_{200}$ is calculated using equation (\ref{eq:c200M200}). Green vertical lines are 
the true cluster parameter values as given in Tab.~\ref{tab:simpars} and the red 
vertical lines are the mean of the probability distributions of the parameters.}  \label{fig:c-Mpos}
\end{figure}

\begin{table}
\caption{Summary of the priors on the sampling parameters. Note that $N(\mu ,\sigma)
$ represents a Gaussian probability distribution with mean $\mu$ and standard 
deviation of $\sigma$ and $U(a,b)$ represents a uniform distribution between $a$ and 
$b$. \label{tab:clpriors2}}
\begin{tabular}{@{}ll@{} }
\hline
Parameter       &\qquad \qquad Prior   \\\hline
$x_{\rm c}$ , $y_{\rm c}\qquad$ &\qquad \qquad $N(0 \,\, , \, \,60)\arcsec$  \\
$\log M_{\rm {tot}}(r_{\rm 200})\qquad$ &\qquad \qquad $U(14 \,\, , \, \, 15.8)\,\rm
{M_\odot}$  \\
$f_{\rm {gas}}(r_{\rm 200})\qquad$ &\qquad \qquad $N( 0.13 \,\, , \, \, 0.02)$ \\ \hline
\end{tabular}
\end{table}
Further details of our Bayesian methodology, modelling interferometric SZ 
data, primordial CMB anisotropies, and resolved and unresolved 
radio point-source models are given in Hobson \& Maisinger (2002), Feroz \& Hobson  
(2008) and Feroz et~al. (2009), AMI Consortium: Davies et~al. (2011) and 
AMI Consortium: Olamaie et~al. (2012).
\begin{figure*}
\centerline{\includegraphics[width=8.0cm,clip=]{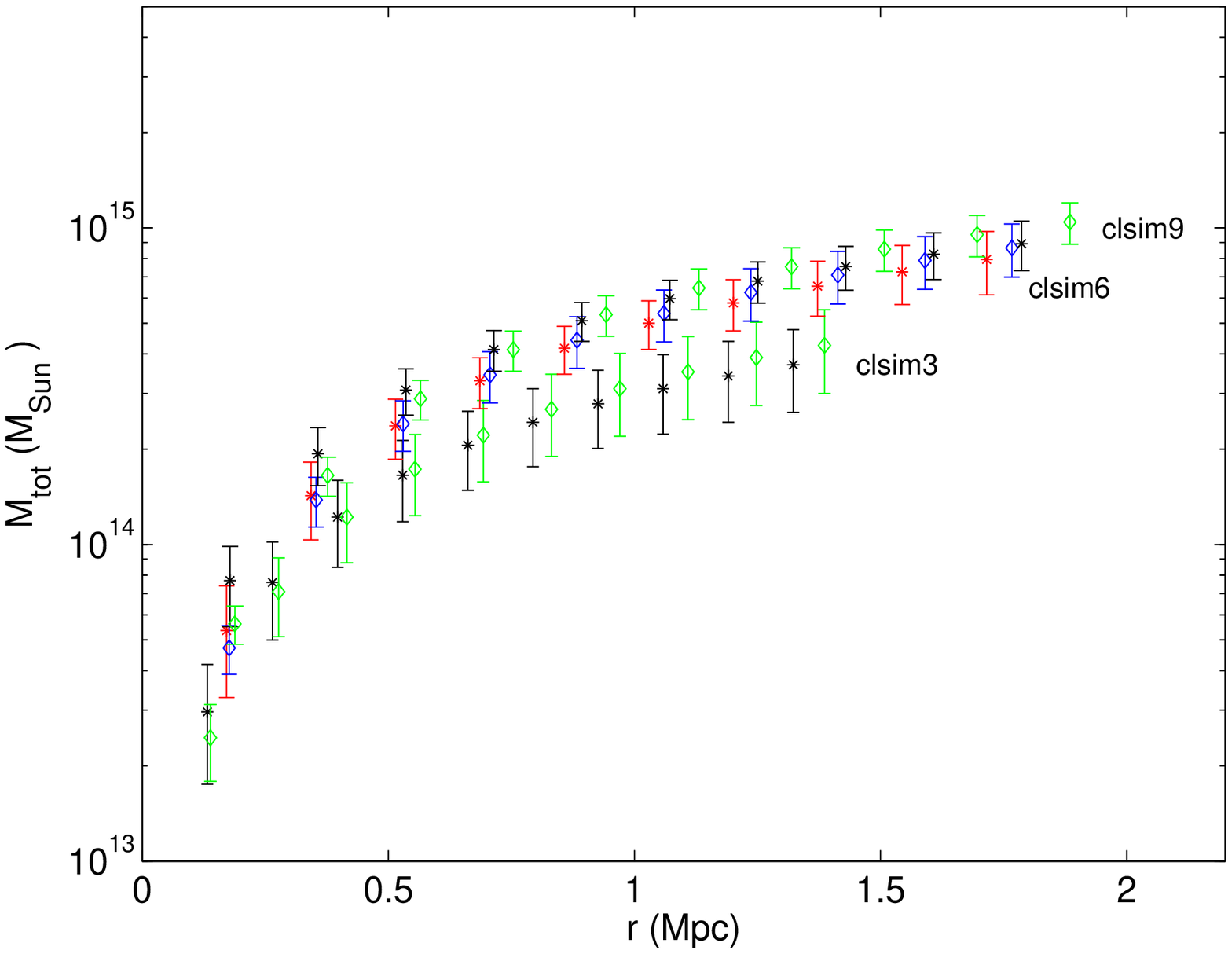} \qquad
            \includegraphics[width=8.0cm,clip=]{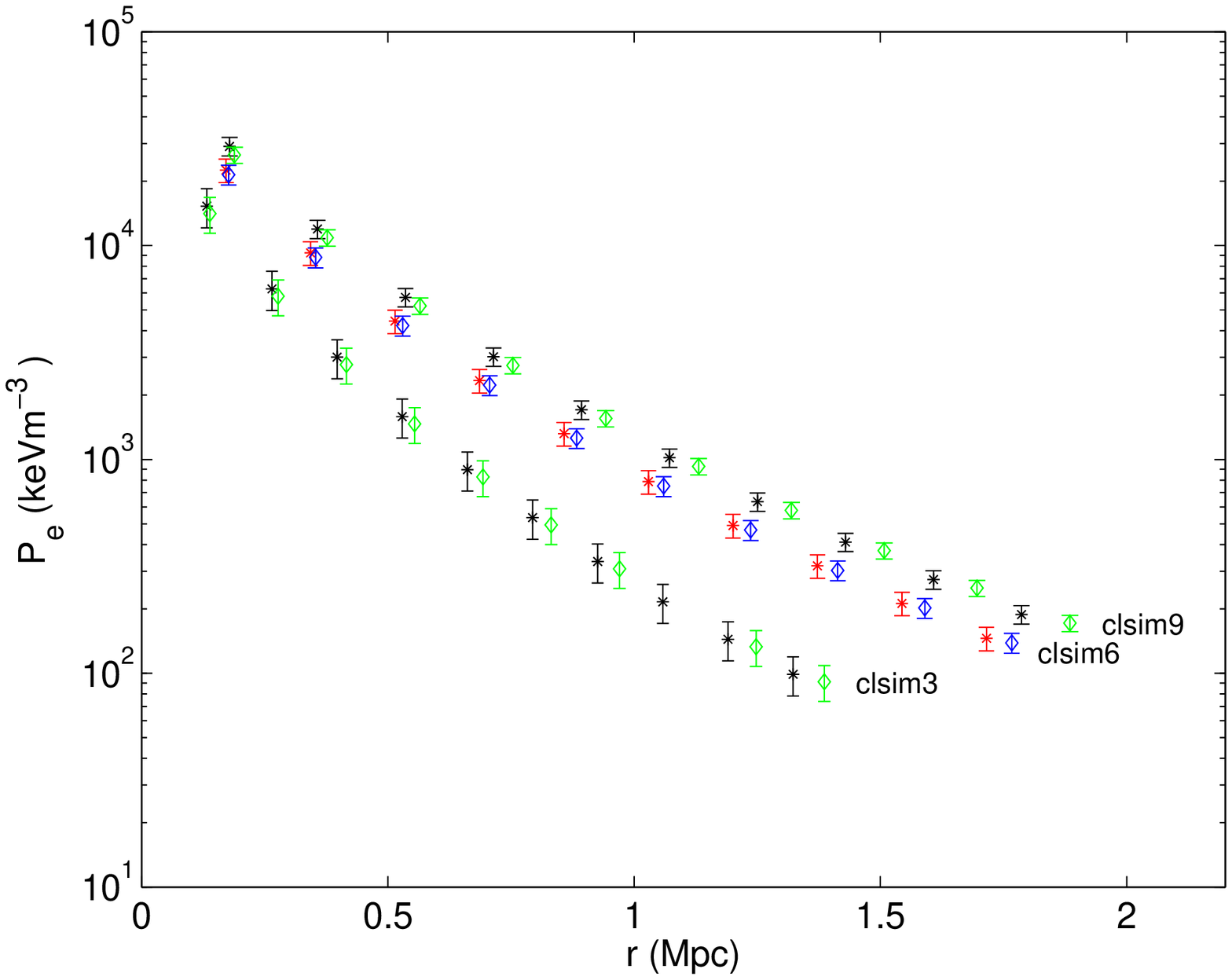} }
\centerline{\includegraphics[width=8.0cm,clip=]{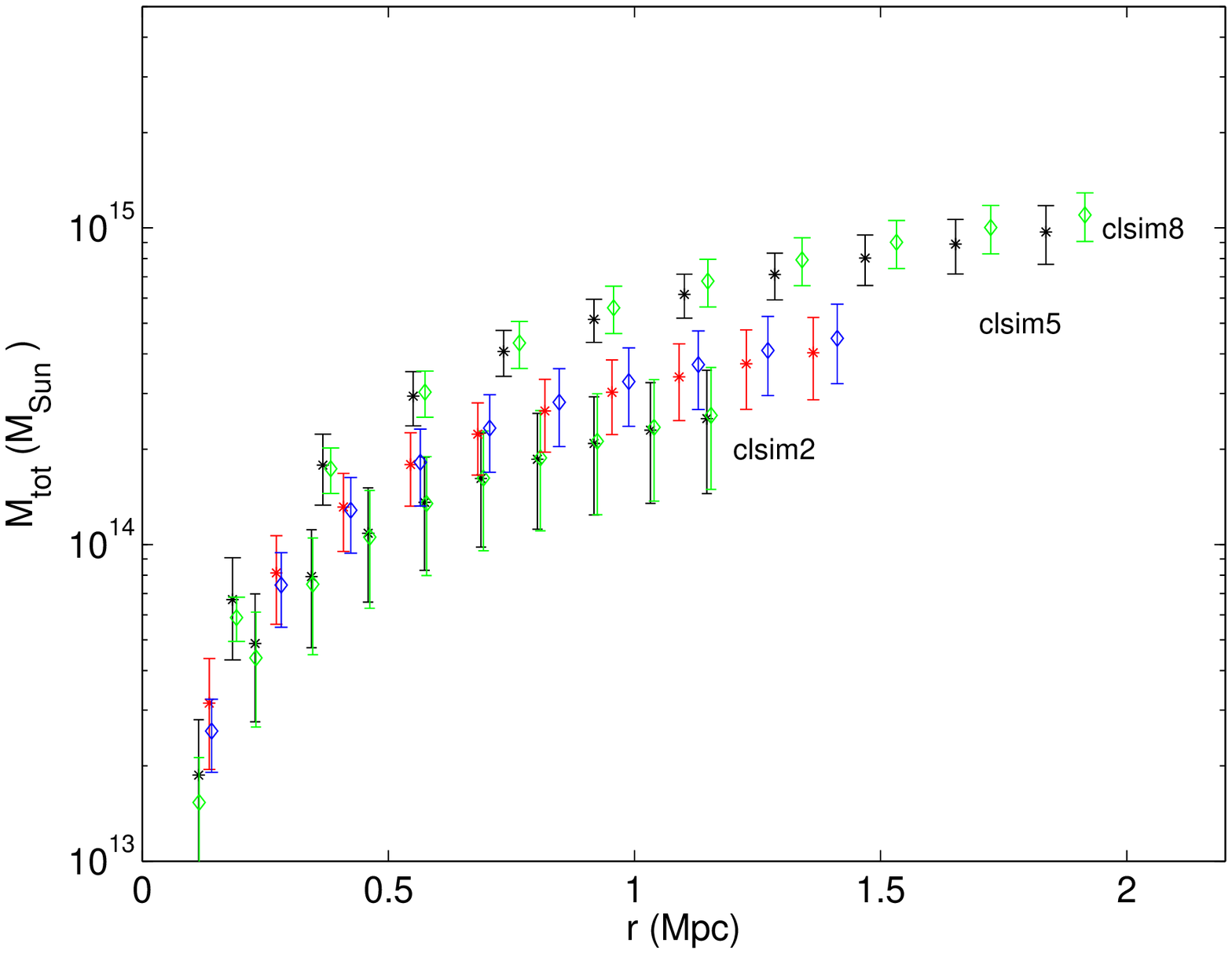} \qquad
            \includegraphics[width=8.0cm,clip=]{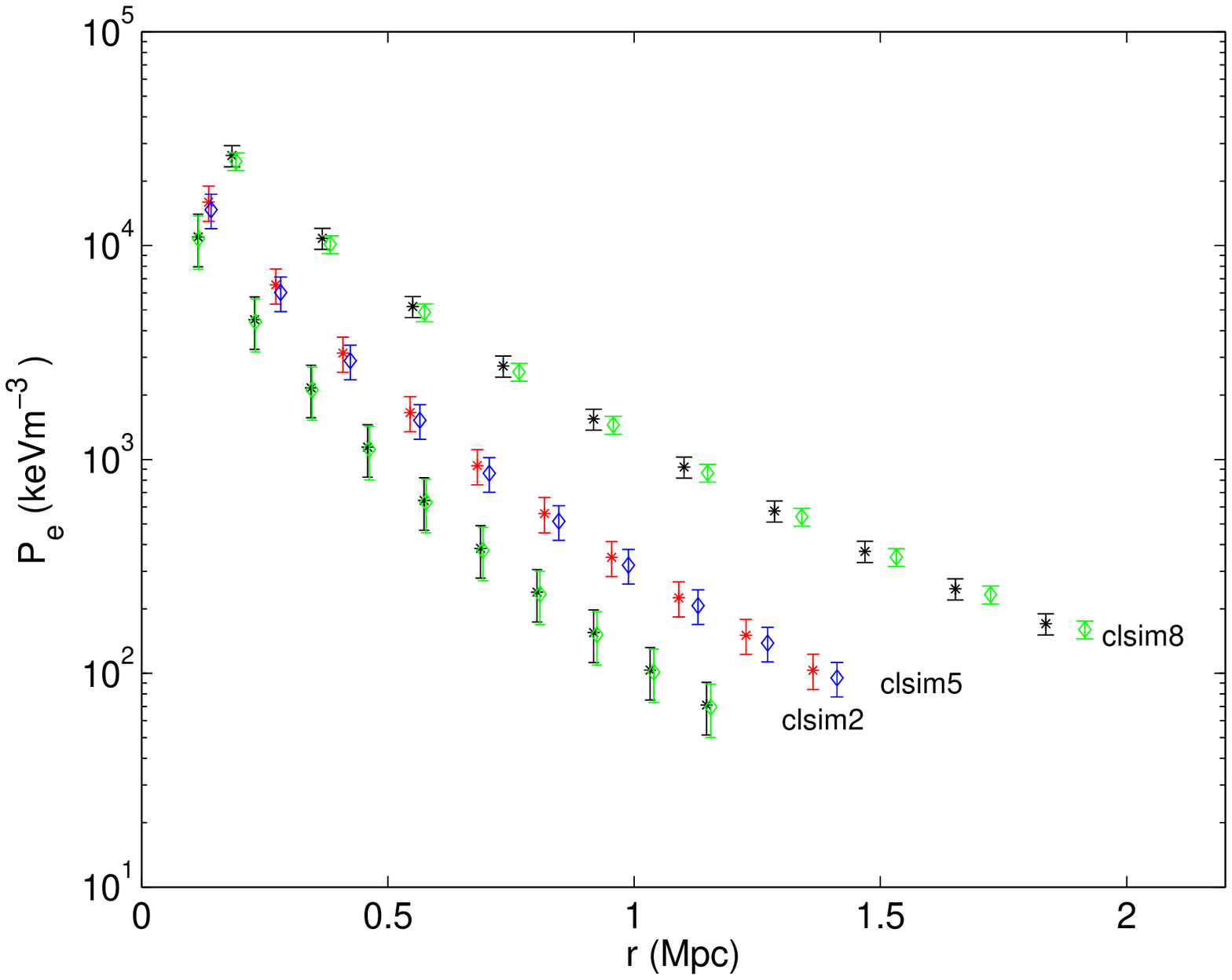}}
\centerline{\includegraphics[width=8.0cm,clip=]{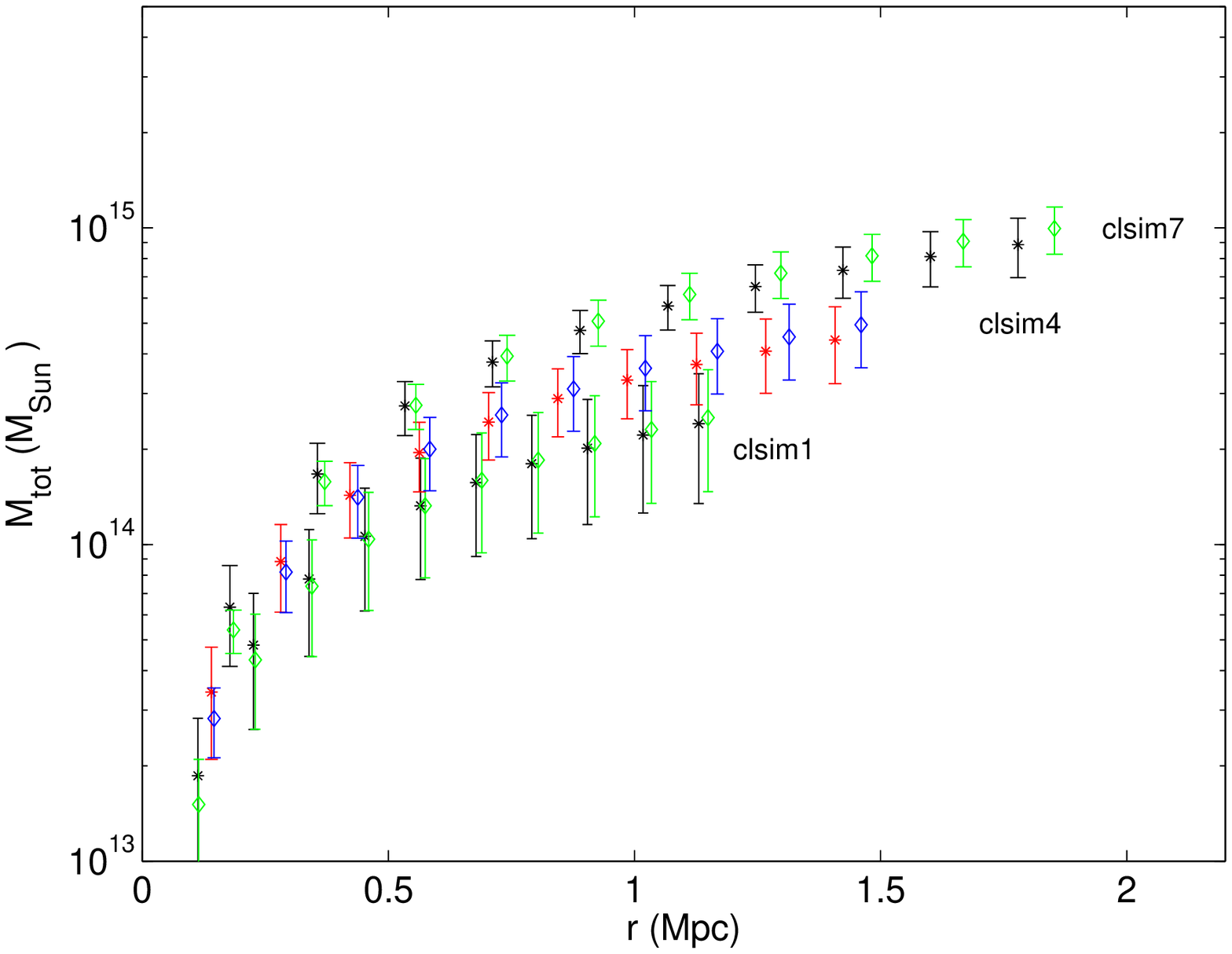} \qquad
            \includegraphics[width=8.0cm,clip=]{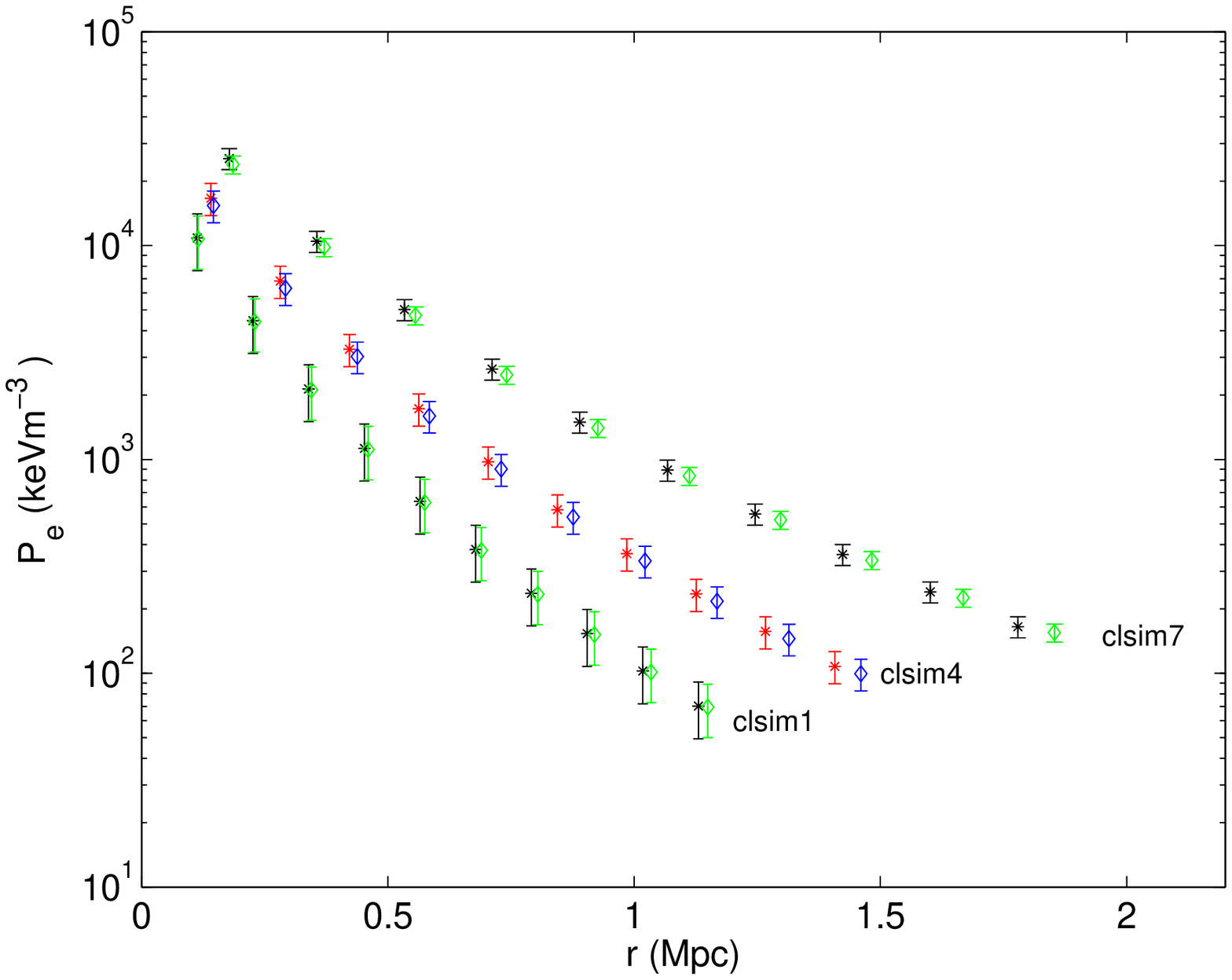}}
\caption{Integrated mass (\emph{left}) and pressure (\emph{right}) profiles as a 
function of $r$ for nine simulated SZ clusters. In each panel, the plots with $\ast$ 
show the profiles when $c_{200}$ is assumed as an input (sampling) parameter and the 
plots with $\diamond$ show the profiles when $c_{200}$ is calculated using equation 
(\ref{eq:c200M200}).\label{fig:m-p-r}} 
\end{figure*}

\section{Results and Discussion} 
Fig.~\ref{fig:varycpos} shows 1D marginalised posterior distributions of sampling 
parameters for clsim2 (one of the lowest mass clusters) when $c_{200}$ is also assumed to be a sampling parameter, 
(Olamaie et~al. 2012). The green vertical lines are the true values of the simulated 
cluster parameters while the red vertical lines represent the mean values of the 
probability distributions. It should be pointed out that similar results are obtained for all clusters in our sample. This results show that while this form of the 
parameterization can constrain cluster projected position on the sky and $M_{\rm 
{tot}}(r_{\rm 200})$, the halo concentration parameter, $c_{200}$, is 
unconstrained and its mean value is just the mean of the prior range which suggests that 
the mean value of the distribution is strongly driven by the prior. This result, as we 
have seen in  our previous studies of clusters of galaxies (AMI 
Consortium: Olamaie et~al. 2012), also suggests a correlation between $c_
{200}$ and the other two physical sampling parameters, i.e. $M_{\rm {tot}}(r_{\rm 200})
$ and $z$. The existence of such a correlation means that the analysis may result in a 
biased estimate of cluster parameters if it is not considered in the analysis.

Fig.~\ref{fig:c-Mpos}  shows 1D marginalised posterior distributions of sampling 
parameters for clsim2 when we take into account the dependency of halo concentration  
on both the formation time and the dynamical state of the halo using equation 
(\ref{eq:c200M200}). From the results it is clear that this form of the 
parameterization can constrain $c_{200}$ as well as other cluster parameters. Moreover, while NFW 
profiles are usually fitted using the two parameters $R_{\rm {s}}$ and $c_{200}$  this 
parametrisation makes the profile a one-parameter profile. We notice that similar results were obtained upon analysing all the clusters in our sample. We also note that $f_{\rm g}(r_{\rm 200})$ is hardly constrained in both parameterisations indicating that the gas fraction can not be constrained using SZ only data. 

The \emph{left} panel in fig.~\ref{fig:m-p-r} presents the integrated mass profiles 
for our sample of nine SZ simulated galaxy clusters and the \emph{right} panel shows 
the pressure profiles of these clusters. In each panel we have plotted the profiles 
out to $r_{200}$ using the two forms of parameterizations: ($1$) assuming $c_{200}$ as 
an input parameter in the analysis ($\ast$) and ($2$) calculating $c_{200}$ using 
equation (\ref{eq:c200M200}), ($\diamond$). 

From the plots, the difference in estimating the cluster mass and its gas pressure 
using two forms of parameterizations is clear and becomes more significant as $M_{\rm 
{tot}}(r_{\rm 200})$ increases. Our resulting constraints on $c_{200}$ indicate that 
because this parameter is completely unconstrained when assumed as a sampling 
parameter, its best fit value is always the mean of the assumed prior range. This 
strong dependency on the prior range may indeed lead to a biased estimate of the 
cluster parameters including its mass.

\section{Conclusion}
We have studied the recovery of $M_{\rm {tot}}(r)$ and $P_{\rm e}(r)$ from the SZ effect for a sample 
of nine simulated galaxy clusters ($2.0 \times 10^{14}\,\rm{M_\odot}<M_{\rm {tot}}(r_
{\rm 200})<1.0\times10^{15}\rm{M_\odot}$) using the model described in Olamaie 
et~al. (2012). This is motivated by the fact that SZ surface brightness is 
proportional to the line of sight integral of the ICM plasma so that SZ data can 
potentially constrain the cluster total mass. 

To obtain an unbiased mass estimate we have carried out a detailed analysis of a 
series of simulated clusters using two different parameterizations within our model 
and its corresponding assumptions (Olamaie et~al. 2012). In the first parameterization 
we assume that the halo concentration parameter $c_{200}$ is also a sampling 
parameter. However, the results of the analysis show that the simulated SZ data can 
not constrain this parameter and therefore its mean value is driven by the prior range 
which may introduce biases in the ultimate cluster mass estimate. This results also 
suggests that $c_{200}$ depends on the other model parameters.

In the second parameterization we consider the correlation of $c_{200}$ with 
$M_{\rm {tot}}(r_{\rm 200})$ and $z$ within $\Lambda \rm{CDM}$ Universe (equation \ref
{eq:c200M200}) as higher mass halos that are forming today are less concentrated than 
halos of lower mass that built up at an earlier epoch, where the mean density was 
higher. This parameterization clearly constrains $c_{200}$ as AMI SZ data can 
constrain $M_{\rm {tot}}(r_{\rm 200})$. We hence conclude that in order to obtain a 
robust estimate on cluster physical parameters including its mass it is crucial to 
consider the mass and redshift dependency of $c_{200}$ as precisely as possible. 
%
%-----------------------------------------------------------
\section*{Acknowledgments}
The authors thank their colleagues in the AMI Consortium for numerous
illuminating discussions regarding the modelling of galaxy clusters. The data 
analyses were carried out on the COSMOS UK National Supercomputer at DAMTP, 
University of Cambridge and we are grateful to Andrey Kaliazin for his computing 
assistance. MO acknowledges an STFC studentship.
%------------------------------------------------------------------------------
%------------------------------------------------------------------------------
\setlength{\labelwidth}{0pt} % fix broken mn2e.cls!

\label{lastpage}
\end{document}